\documentclass{article}
\usepackage{graphicx}
\usepackage[margin=1.5in]{geometry}

\begin{document}

\title{Illusory Decoherence}
\author{Sam Kennerly}

\maketitle

\noindent
Email: samkennerly@gmail.com\\
Website: http://sites.google.com/site/samkennerly\\
Institution: Drexel University

\begin{abstract}
\noindent
If a quantum experiment includes random processes, then the results of repeated measurements can appear consistent with irreversible decoherence even if the system's evolution prior to measurement was reversible and unitary. Two thought experiments are constructed as examples.
\end{abstract}

\


\section{Introduction}
\label{intro}

Time evolution according to the Schr\"odinger equation (or equivalent formulations) is deterministic, unitary, and cannot alter von Neumann entropy. Despite this fact, quantum computing experiments routinely produce data which appear to show decoherence of pure states into mixed states. The fickle behavior of qubits lends yet more support to the widely-used principle that physical systems tend irreversibly toward disorder. As summarized by Eddington:
\begin{quote}
The law that entropy always increases holds, I think, the supreme position among the laws of Nature... If your theory is found to be against the second law of thermodynamics I can give you no hope; there is nothing for it but to collapse in deepest humiliation.\cite{Eddington_1927}
\end{quote}

The Schr\"odinger equation has not yet collapsed in deepest humilation. But its apparent conflict with the second law is not easily dismissed: how can a reversible theory produce irreversible evolution? This \emph{quantum Loschmidt paradox} (or \emph{reversibility paradox}) is essentially a modern reformulation of Loschmidt's criticism of Boltzmann's H-theorem.\footnote{The quantum Loschmidt paradox should not be confused with the cosmological time-reversal paradox, which may or may not be related. See e.g. \cite{Carroll_book} for a discussion of both.} If time evolution of quantum systems is unitary, then von Neumann entropy does not tend to increase. Does von Neumann entropy disobey the second law, or do states evolve in a non-unitary way?

The following thought experiments are examples for which the answer is ``none of the above.'' In each experiment, a quantum Loschmidt paradox is created by careless use of the term ``entropy.'' The paradoxes are resolved by defining von Neumann entropy exclusively for \emph{statistical mixtures}, not for \emph{physical objects}.

\section{Allyson's choice}
\label{AC}

\begin{figure}
\begin{center}
\includegraphics[width=0.7\textwidth]{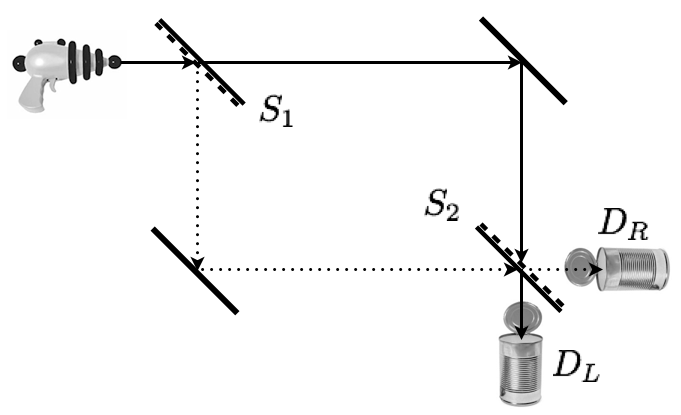}
\caption{An idealized Mach-Zehnder inferferometer. A single neutron is sent through splitter $S_1$. The resulting superposition of paths is reflected from mirrors, altered by splitter $S_2$, and sent to detectors $D_L, D_R$. Solid lines label path $|L\rangle$ and dotted lines label path $|R\rangle$.\label{fig:MZA}
}
\end{center}
\end{figure}

Professor Bob intends to replicate a classic \emph{welcher-weg} experiment for his students. On each trial, Bob sends a single neutron through a Mach-Zehnder interferometer as shown in Figure \ref{fig:MZA}. Bob performs 1000 such trials, adjusts the mirrors so that the phase difference $\phi$ between the paths is increased to $\phi + \Delta$, then repeats this procedure many times. When Bob plots his detector counts as a function of $\phi$, he expects to see sinusoidal $\phi$ dependence due to de Broglie interference.

As a practical joke, Bob's student Allyson subverts the experiment. Before each trial, she flips a fair coin and records the result.\footnote{To avoid the difficulty of flipping thousands of coins without attracting the suspicion of her advisor, suppose she automates this process with a quantum RNG instead of a coin.} If heads, she performs the experiment as planned. If tails, she covertly reverses the orientation of the second beamsplitter. It is then very probable that Bob's plot of detector counts will show almost no $\phi$ dependence. Though it appears to Bob that quantum information has been destroyed, it has actually been encrypted. Using her coin-flip history as a password, Allyson can decrypt Bob's data to produce two plots, each of which will show the sinusoidal dependence predicted by quantum mechanics.

In each of these examples, the neutron is assumed to evolve unitarily on each trial. De Broglie interferometry is used because it is a well-documented topic that is relatively easy to visualize. Neutrons are chosen to avoid questions of relativity and electrodynamics, but other uncharged, massive particles would be suitable as well.\footnote{Similar experiments with different interferometer confgurations have been performed with C$_{60}$ ``buckyballs'' and even larger molecules.\cite{Buckyball}\cite{Large_molecules}} For realizations of such an experiment, see e.g.\ \cite{MZ_neutron}\cite{MZ_sodium}.

\subsection{Bob's intended experiment}
\label{AC_Bob}

A neutron is sent through the Mach-Zehnder apparatus shown in Figure \ref{fig:MZA}. The beamsplitter $S_1$ sends the neutron to a superposition of paths $|L\rangle$ and $|R\rangle$. Mirrors send the paths to a second splitter $S_2$, then to detectors $D_L$ and $D_R$. If the splitters are of the lossless type described in \cite{Z_beamsplitter}, then each can be represented by a unitary matrix acting on a Hilbert space with basis $\{ |L\rangle, |R\rangle \}$:
$$
\hat{S} =
\left[\begin{array}{cc}
r_{LL} & t_{LR} \\
t_{RL} & r_{RR}
\end{array}\right]
\qquad
r_{RR} = r_{LL}^*,
\quad
t_{RL} = -t_{LR}^*,
\quad
\det[\hat{S}] = 1
$$
The parameters $r_{LL}, r_{RR}, t_{LR}, t_{RL}$ are (complex) reflection and transmission coefficients. Bob chooses splitters $S_1$ and $S_2$ as follows:
$$
\hat{S}_1 =
\frac{1}{\sqrt{2}}
\left[\begin{array}{cc}
1 & 1 \\
-1 & 1
\end{array}\right]
\qquad
\hat{S}_2 = (\hat{S}_1)^{-1} =
\frac{1}{\sqrt{2}}
\left[\begin{array}{cc}
1 & -1 \\
1 & 1
\end{array}\right]
$$
Represent the phase shifts of the neutron's wavefunction along the two paths by an operator $\hat{\Phi}$. The combined action of the interferometer is then $\hat{H} \equiv \hat{S}_2\hat{\Phi}\hat{S}_1$.\footnote{The notation $\hat{H}$ is chosen to suggest ``heads,'' not ``Hamiltonian.''}
$$
\hat{\Phi} = 
\left[\begin{array}{cc}
e^{\imath \theta_L} & 0 \\
0 & e^{\imath \theta_R}
\end{array}\right]
\quad
\hat{H} = 
\hat{S}_2\hat{\Phi}\hat{S}_1 =
e^{\imath \theta_L} \frac{1}{2}
\left[\begin{array}{cc}
1 + e^{\imath \phi}  & 1 - e^{\imath \phi}  \\
1 - e^{\imath \phi}  & 1 + e^{\imath \phi}
\end{array}\right]
\quad
\phi \equiv \theta_R - \theta_L
$$
If the neutron is launched as shown in Figure \ref{fig:MZA}, then it reaches the detectors in state $|\Psi_H\rangle \equiv \hat{H}| L \rangle$. Ignoring any unobservable overall phase, $|\Psi_H\rangle$ is:
$$
|\Psi_H\rangle
\equiv
\hat{H}| L \rangle
=
\frac{1}{2}
\left[\begin{array}{cc}
1 + e^{\imath \phi}  & 1 - e^{\imath \phi}  \\
1 - e^{\imath \phi}  & 1 + e^{\imath \phi}
\end{array}\right]
\left[\begin{array}{c}
1 \\
0
\end{array}\right]
=
\frac{1}{2}
\left[\begin{array}{c}
1 + e^{\imath \phi} \\
1 - e^{\imath \phi}
\end{array}\right]
$$
The detection probabilities $P(D_L)$ and $P(D_R)$ are:
\begin{center}$
P(D_L) = 
|| \langle L | \Psi_H \rangle ||^2
= || \frac{1}{2}(1 + e^{\imath \phi})||^2
= \frac{1}{2}( 1 + \cos \phi )
$\end{center}
\begin{center}$
P(D_R) = 
|| \langle R | \Psi_H \rangle ||^2
= || \frac{1}{2}(1 - e^{\imath \phi}) ||^2
= \frac{1}{2} ( 1 - \cos \phi )
$\end{center}

\subsection{Allyson's randomized experiment}
\label{AC_Allyson}

\begin{figure}
\begin{center}
\includegraphics[width=0.4\textwidth]{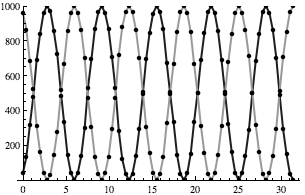}
\qquad
\includegraphics[width=0.4\textwidth]{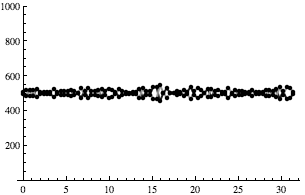}
\caption{Simulated data with 1000 trials for each value of $\phi$. Y-axis is detector counts; X-axis is $\phi$. Left: Bob's intended experiment. Right: Allyson's randomized experiment. (Imperfect fit to predicted probabilities is an artifact of finite sample size.)\label{fig:Bob_fail}}
\end{center}
\end{figure}

When Allyson's coin lands heads, the neutron state immediately prior to detection is $\hat{H}| L \rangle$. When it lands tails, she reverses the orientation of $S_2$ so that its matrix representation is $(\hat{S}_2)^T = \hat{S}_1$ and the action of the M-Z apparatus is $\hat{T} \equiv \hat{S}_1 \hat{\Phi} \hat{S}_1$. For these trials, the state vector immediately before detection is:
$$
|\Psi_T\rangle
\equiv \hat{T}|L \rangle
=
\frac{1}{2}
\left[\begin{array}{cc}
1 - e^{\imath \phi}  & 1 + e^{\imath \phi}  \\
- (1 + e^{\imath \phi})  & -1 + e^{\imath \phi}
\end{array}\right]
\left[\begin{array}{c}
1 \\
0
\end{array}\right]
=
\frac{1}{2}
\left[\begin{array}{c}
1 - e^{\imath \phi} \\
- (1 + e^{\imath \phi})
\end{array}\right]
$$
Given tails, the conditional probabilities $P(D_L | T)$ and $P(D_R | T)$ are:
\begin{center}$
P( D_L | T)
= || \langle L | \Psi_T \rangle ||^2
= || \frac{1}{2}(1 - e^{\imath \phi})||^2
= \frac{1}{2}( 1 - \cos \phi )
$\end{center}
\begin{center}$
P( D_R | T)
= || \langle R | \Psi_T \rangle ||^2
= || \frac{-1}{2}(1 + e^{\imath \phi})||^2
= \frac{1}{2}( 1 + \cos \phi )
$\end{center}
Given heads, the conditional probabilities $P(D_L | H)$ and $P(D_R | H)$ are the same as in Bob's intended experiment. The unconditional probabilities are:
$$
P(D_L) = P(D_L | H) P(H) + P(D_L | T) P(T) = \frac{1}{2}
$$
and likewise for $P(D_R)$, which is also $\frac{1}{2}$ regardless of the value of $\phi$. Figure \ref{fig:Bob_fail} shows numerical simulations of the intended and randomized experiments.

Allyson's shenanigans have concealed evidence of de Broglie interference, and Bob's experiment appears ruined. However, Allyson can unscramble the data if she has recorded her coin tosses. Suppose the first 8 coin toss results were $THHT$ $THTH$. Then Allyson represents the first 8 final states as a list:
$$
\begin{array}{ccccccccc}
|\Psi_T\rangle, & |\Psi_H\rangle, & |\Psi_H\rangle, & |\Psi_T\rangle & \quad &
|\Psi_T\rangle, & |\Psi_H\rangle, & |\Psi_T\rangle, & |\Psi_H\rangle \\
\end{array}
$$
By contrast, Bob's incorrect final-state list contains exclusively $|\Psi_H \rangle$ entries.

\begin{table}
\begin{center}
\caption{Example final state list (first 8 trials only)\label{tab:state_lists}}
\begin{tabular}{ccccccccc}
\hline\noalign{\smallskip}
Bob's plaintext & 1 & 1 & 1 & 1 & 1  & 1 & 1 & 1 \\
\noalign{\smallskip}\hline\noalign{\smallskip}
Coin result & T & H & H & T & T & H & T & H \\
\noalign{\smallskip}\hline\noalign{\smallskip}
Allyson's ciphertext & 0 & 1 & 1 & 0 & 0 & 1 & 0 & 1 \\
\noalign{\smallskip}\hline
\end{tabular}
\end{center}
\end{table}

Consider each physicist's list as a binary string with $|\Psi_H \rangle \sim 1$ and $|\Psi_T \rangle \sim 0$. Then Bob's list is a plaintext, Allyson's list is a ciphertext, and her coin history is a password. The encryption scheme is bitwise modular binary addition as shown in Table \ref{tab:state_lists}. Allyson's password is the same size as the plaintext, used only once, and chosen randomly with uniform probability. Her encryption scheme is thus a provably-secure Vernam cipher.\cite{Shannon_49} Even if Bob discovers Allyson's subterfuge, he cannot decrypt the data unless he knows the results of her coin flips.

\begin{figure}
\begin{center}
\includegraphics[width=0.4\textwidth]{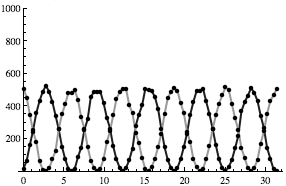}
\qquad
\includegraphics[width=0.4\textwidth]{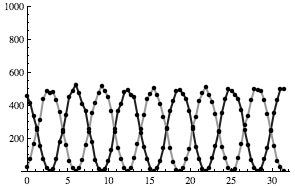}
\caption{Decryption of simulated data from Figure \ref{fig:Bob_fail}. Y-axis is detector counts; X-axis is $\phi$. Left: Heads trials. Right: Tails trials.\label{fig:Allyson_decrypt}}
\end{center}
\end{figure}

Allyson can decrypt the data by using her coin-toss history to label each detection event H or T. She can then produce separate heads-only and tails-only plots as shown in Figure \ref{fig:Allyson_decrypt}. These plots show that evidence of the neutrons' de Broglie interference was reversibly \emph{encrypted}, not irreversibly \emph{destroyed}.

\subsection{Mixed-state description of the experiment}
\label{AC_mixed}

Suppose Allyson recorded her coin history on a flash memory stick, which she has now misplaced. Unless she made backup copies or kept some other record of coin flips, the password is lost and neither Allyson nor Bob can decrypt the data.

Decryption without the password is extremely unlikely, but a mixed-state representation of the neutrons' final state remains possible. Allyson and Bob know only that the final state on each trial might have been $|\Psi_H\rangle$ or $|\Psi_T\rangle$, each with probability $\frac{1}{2}$. The projection operators $\hat{\rho}_H, \hat{\rho}_T$ corresponding to these states are:
$$
\hat{\rho}_H =
|\Psi_H \rangle \langle \Psi_H | =
\frac{1}{2}
\left[\begin{array}{cc}
1 + \cos(\phi) & \imath \sin(\phi) \\
- \imath \sin(\phi) & 1 - \cos(\phi)
\end{array}\right]
$$
$$
\hat{\rho}_T =
|\Psi_T \rangle \langle \Psi_T | =
\frac{1}{2}
\left[\begin{array}{cc}
1 - \cos(\phi) & \imath \sin(\phi) \\
- \imath \sin(\phi) & 1 + \cos(\phi)
\end{array}\right]
$$
Following von Neumann's prescription for ``when we do not even know what state is actually present,'' Allyson and Bob weight each of these projections by its probability and sum the results to form a mixed state $\bar{\rho}$.\cite{vN_foundations}
$$
\bar{\rho} =
\frac{1}{2}\hat{\rho}_H + \frac{1}{2}\hat{\rho}_T =
\frac{1}{2}
\left[\begin{array}{cc}
1 & \imath \sin(\phi) \\
- \imath \sin(\phi) & 1
\end{array}\right]
$$
This $\bar{\rho}$ predicts detection probabilities of $\frac{1}{2}$ for all values of $\phi$, agreeing with Bob's observations. Its eigenvalues $\lambda_{\pm}$ and von Neumann entropy $S_{vN}$ are:
$$
S_{vN} = - \sum \lambda_{\pm} \log_2(\lambda_{\pm})
\qquad
\lambda_{\pm} = \frac{1}{2}\left[ 1 \pm \sin(\phi) \right]
$$
This entropy varies smoothly from to 1 bit to 0 bits depending on the value of $\phi$ for a particular trial. When $\sin(\phi) = 1$, the neutron is equally likely to be detected by $D_L$ or $D_R$ regardless of Allyson's coin flip. But when $\sin(\phi) = 0$, her subterfuge transforms what should have been a certain event into a 50/50 proposition.

\section{Decoherence by 1000 small cuts}
\label{1000_cuts}

Suppose Bob repeats the experiment with Allyson's cooperation. Suppose also that this time, their control of the interferometer is imperfect in a specific way: the phase difference $\phi$ along the neutron paths varies erratically by an amount that is not negligible but is impractical to measure directly.\footnote{The physical source of imprecision in $\phi$ is left to readers' imaginations; perhaps it is seismic vibrations, flexibility of the beamsplitters' mounting brackets, or some other nuisance.}

Bob now faces a subtler version of his previous difficulty. For each value of $\phi$, he assumes that performing 1000 trials will ensure the ratio of $D_L/D_R$ detections is close to its expectation value $||\langle L | \Psi \rangle ||^2 \ / \ || \langle R | \Psi \rangle ||^2$. In the idealized experiment, this expectation is identical over 1000 trials and Bob's assumption follows from the law of large numbers.\footnote{Bob must also assume that detection events for different trials are independent.} But imprecision in $\phi$ means $|\Psi\rangle$ is not identical for all 1000 trials, which invalidates Bob's reasoning.

Errors in $\phi$ prevent either physicist from knowing $|\Psi\rangle$ exactly on each trial. If instead they represent $\phi$ as a random variable which is identically \emph{distributed} over 1000 trials, then they can describe the neutrons' final state as a statistical mixture. For simplicity, let $\phi$ be normally-distributed with mean $\mu$ and variance $\sigma^2$.\footnote{If $\phi$ is the sum of very many independent random variables with finite mean and variance, then this assumption is justified by the central limit theorem.} Assume also that $\sigma > 0$ is fixed, but the experimenters' control of $\mu$ is nearly perfect. For each choice of $\mu$, they define $\bar{\rho}(\mu)$ as a conditional expectation:
$$
\bar{\rho}(\mu)
\equiv
E(\hat{\rho} | \mu )
= \frac{1}{\sigma \sqrt{2 \pi}}
\int_{-\infty}^{\infty}
\frac{1}{2}
\left[\begin{array}{cc}
1 + \cos(\phi) & \imath \sin(\phi) \\
- \imath \sin(\phi) & 1 - \cos(\phi)
\end{array}\right]
e^{-\frac{1}{2}\left( \frac{\phi - \mu}{\sigma} \right)^2}
\ d \phi
$$
Each matrix element is then a convolution. The resulting mixture has $S_{vN} > 0$:
$$
\bar{\rho}(\mu)
= \frac{1}{2}
\left(
\left[\begin{array}{cc}
1 & 0 \\
0 & 1
\end{array}\right]
+ \left[\begin{array}{cc}
\cos(\mu) & \imath \sin(\mu) \\
-\imath \sin(\mu) & -\cos(\mu)
\end{array}\right]
e^{-\frac{1}{2}\sigma^2}
\right)
$$
If $\sigma^2 \ll 1$, then $\bar{\rho}$ is nearly pure and a plot of detection counts versus $\mu$ is likely to closely match Bob's intentions. As $\sigma^2$ increases, the amplitude of $\mu$-dependence decreases exponentially. If $\sigma^2 \gg 1$, then $\bar{\rho}$ approaches the maximum-entropy mixed state and Bob's plot shows no evidence of de Broglie interference. 

In principle, a similar analysis can be applied to any controllable two-level quantum system. For example, the operators $\hat{S}_1, \hat{\Phi}, \hat{S}_2$ could represent transformations of a superconducting qubit state as performed in \cite{MZ_qubit} or \cite{NMR_control}. The same mathematical formalism can be used, though different physical sources of experimental errors may require different noise models, e.g. \cite{Bias_noise} or \cite{Dielectric_loss}.

The conclusion that noisy experiments can produce decoherence is unlikely to surprise many experimental physicists. However, it may be surprising that any credible prediction can be made from such a crudely simplified model. There was no attempt to describe the laboratory environment or any extra degrees of freedom for the neutron - only an assumption that on each trial, $\phi$ is a random variable which is i.i.d.\ normal with mean $\mu$ controlled by experimenters.

\section{Interpretation and conclusions}
\label{interpretation}

For Allyson's randomized experiment, Bob's quantum Loschmidt paradox is: ``How did unitary evolution of a pure state produce data consistent with decoherence?'' Allyson's resolution is: ``Evolution was unitary during each trial but \emph{the trials were not identical}.'' The classical appearance of Bob's data is due to his erroneous assumption that the neutron evolved identically for each batch of 1000 trials.

In the ``1000 cuts'' experiment, the paradox and resolution are the same: evolution was unitary, but the trials were not identical. In this case, encryption was performed by \emph{Bob's laboratory} rather than a mischievous graduate student, and the password is the precise history of errors in $\phi$. Without knowing the password, Allyson and Bob cannot exactly represent $|\Psi\rangle$ on each trial. At best, they can resort to a probabilistic representation in terms of a density matrix $\bar{\rho}$. This $\bar{\rho}$ evolves irreversibly with $\Delta S_{vN} > 0$ while the neutrons themselves evolve unitarily.

\subsection{Where did the information go?}
\label{missing_information}

Shannon interpreted the quantity $- \sum p_n \log(p_n)$ as a measure of ``missing information.''\cite{Shannon_48} That interpretation can be taken quite literally in these examples. Because any mixture $\bar{\rho}$ is a convex combination of projection operators, it can be used to define a probability distribution of pure states. Given a mixture $\bar{\rho}$, diagonalize it and refer to its eigenvectors $\{ | \Psi_n \rangle \}$ as ``possible pure states.''\footnote{Diagonalization does not determine the overall phase of each eigenvector, but these phases are arbitrary and do not represent any physically observable quantity.} The corresponding eigenvalues $\{ \lambda_n \}$ then form a probability distribution in the usual sense: $\lambda_n \in [0,1]$ and $\sum \lambda_n = \textrm{Tr}[\bar{\rho}] = 1$. The von Neumann entropy of $\bar{\rho}$ is the Shannon entropy of its associated probability distribution.

In Allyson's randomized experiment, the coin-toss history is needed to determine whether $|\Psi_H\rangle$ or $|\Psi_T\rangle$ occurred on each trial. That information was missing from Bob's record of the experiment, so Allyson could represent the neutrons with a sequence of pure states and Bob could not. When Allyson then misplaces her flash memory, it becomes ``missing information'' from her point of view as well. In the other example, the ``missing'' history of errors in $\phi$ is not misplaced or hidden; it was simply never recorded. In both cases information has not been irreversibly destroyed or removed from the universe. It is ``missing'' only in the sense that neither physicist has any practical means of recovering it. 

It should be emphasized that ``missing information'' here refers to data needed to specify a \emph{state vector}, not a \emph{measurement result}. Given a flawless record of the neutron's evolution, Allyson and Bob can predict a unique $|\Psi \rangle$ on any given trial. But even in an idealized noiseless experiment, quantum theory asserts that neither physicist can predict which detector will detect the neutron unless $\sin(\phi) = 0$. If so, then the information needed to predict specific measurement results is fundamentally inaccessible, not merely ``missing.''\footnote{Unorthodox theories (e.g. Bohmian mechanics or the stochastic-spacetime interpretation) may consider this information accessible in principle but missing from quantum theory.}

\subsection{Entropy of mixtures versus entropy of objects}
\label{entropy_of_what}

In these experiments, it is unclear how to answer the question ``What is the entropy of the neutron?'' If Allyson knows the neutron's exact history and Bob does not, then the ``entropy of the neutron'' appears to be zero for Allyson and nonzero for Bob. By contrast, the question ``What is the von Neumann entropy of the mixture $\bar{\rho}$?'' has a unique answer. The distinction is semantic, but important: $S_{vN}$ is well-defined for \emph{statistical mixtures}, not for \emph{physical objects}.

A classical analogy may be more intuitive. Suppose Bob shuffles a new deck of 52 cards while Allyson films with a high-speed camera. By replaying the shuffle in slow motion, she determines each card's new location and concludes that the entropy of the deck is zero. If Bob has not seen the video, then he concludes that the entropy of the deck is $\log(52!)$. This ambiguity can be avoided by defining Shannon entropy exclusively for probability distributions. Allyson has assigned a degenerate distribution to the set of all deck permutations, while Bob has assigned a uniform distribution.\footnote{A discrete distribution is \emph{degenerate} iff its support consists of exactly one value.} These distributions have well-defined Shannon entropies even if the deck of cards itself does not. Shuffling did not irreversibly alter the deck of cards -- it merely obscured Bob's knowledge of the cards' order.

A similar distinction between entropy of \emph{objects} and entropy of \emph{experiments} was advocated by Jaynes in 1957:
\begin{quote}
It is possible to maintain the view that the system is at all times in some definite but unknown pure state, which changes because of definite but unknown external forces; the probabilities represent only our ignorance as to the true state. With such an interpretation the expression ``irreversible process'' represents a semantic confusion; it is not the physical process that is irreversible, but rather our ability to follow it.\cite{Jaynes_1957_II}
\end{quote}
Jaynes' statement was made in the context of semiclassical statistical mechanics, but it is also relevant here. By assumption, each trial produces a pure final state $|\Psi\rangle$. In these thought experiments, it is the physicists' ability to describe $| \Psi \rangle$ which evolves irreversibly, not the neutrons themselves.

\subsection{Relation to Jaynes' subjective statistical mechanics}
\label{subjectivity}

The interpretation of $S_{vN}$ advocated here can be summarized as follows:
\begin{quote}
$S_{vN}$ is a measure of the missing information an experimenter needs in order to distinguish a pure state $|\Psi\rangle$ from a statistical mixture $\bar{\rho}$.
\end{quote}
According to this interpretation, $S_{vN}$ is ``anthropomorphic'' in the sense that it is a measure of a scientist's inability to precisely represent a physical system, not a natural property of the system itself. Jaynes made the stronger statement (which he attributed to Wigner) that \emph{all} entropy is anthropomorphic:
\begin{quote}
Entropy is an anthropomorphic concept, not only in the well-known statistical sense that it measures the extent of human ignorance as to the microstate. Even at the purely phenomenological level, entropy is an anthropomorphic concept. For it is a property, not of the physical system, but of the particular experiments you or I choose to perform on it.\cite{Jaynes_1965}
\end{quote}
The implications of this interpretation are still a topic of active research.\cite{FoundPhys_MaxEnt_2011}

Anthropomorphic entropy appears to be a useful concept for describing experimental quantum decoherence. In particular, the quantum Loschmidt paradox is avoided by defining $S_{vN}$ exclusively for mixtures resulting from random models of imperfectly-controlled experiments. But if \emph{thermodynamic} entropy $S_T$ is also defined anthropomorphically, then one must be careful to avoid subjectivity paradoxes.\footnote{Jaynes' original papers on subjective statistical mechanics address this issue.\cite{Jaynes_1957_I}} In the randomized experiment, the value Bob calculates for $S_{vN}$ depends on whether he knows Allyson's coin-toss history. Thermodynamic quantities presumably do not depend on physicists' knowledge; precisely describing the state vector of a boiling pot of water does not prevent it from scalding one's finger.

Whether $S_T$ should also be interpreted anthropomorphically is not directly addressed by the thought experiments described here. None of these examples invokes any thermodynamic laws or definitions, nor is there any assumption of equilibrium with an environment. Consequently the neutrons' thermodynamic entropies $S_T$ need not even be well-defined quantities -- and if they are, there is no reason to assume that $S_T$ is related to $S_{vN}$ in either example. It is thus hard to see how these thought experiments could support or refute any statements about thermodynamics. But while they neither support nor refute Jaynes' interpretation, they are consistent with it. Jaynes defined $S_T$ as a special case of $S_{vN}$: it is $S_{vN}$ of the mixture $\bar{\rho}_{max}$ which maximizes entropy for a given macrostate.\footnote{The usual method of MaxEnt quantum thermodynamics is: given a Hilbert space and a set of expectation values $\{ \langle F_i \rangle \}$, define an equilibrium mixture $\bar{\rho}_T$ as the density operator which maximizes $S_{vN} - \sum \lambda_i \langle F_i \rangle$. Here $\{ \langle F_i \rangle \}$ is called a \emph{macrostate} and $\{ \lambda_i \}$ are Lagrange multipliers. The von Neumann entropy of $\bar{\rho}_T$ is then identified with $S_T$ for that macrostate.\cite{Jaynes_1957_II}} For a given macrostate, $\bar{\rho}_{max}$ is not subjective and need not equal Bob's $\bar{\rho}$.

The interpretation of $S_{vN}$ advocated here is consistent with Jaynes' view but is less ambitious. The purpose of these thought experiments is simply to show that unitary evolution can \emph{appear} to produce evolution of pure states into mixed states.

\section{Acknowledgements}
This paper was inspired by discussions of Wheeler's delayed-choice experiment with Prof.\ Robert Gilmore and graduate student Allyson O'Brien at Drexel University.

\bibliography{Illusory_Decoherence}
\bibliographystyle{unsrt}

\end{document}